\documentclass[sigplan,screen,dvipsnames]{acmart}

\usepackage[utf8]{inputenc}

\setcopyright{acmcopyright}
\copyrightyear{2022}
\acmYear{2022}

\acmConference[PERM '22]{PERM '22: ACM SIGPLAN Workshop on Partial Evaluation and Program Manipulation 
}{January 17--18, 2022}{Philadelphia, PA}
\acmBooktitle{PERM '22: ACM SIGPLAN Workshop on Partial Evaluation and Program Manipulation,
January 17--18, 2022, Philadelphia, PA}
\acmPrice{15.00}
\acmISBN{978-1-4503-XXXX-X/18/06}

\setcopyright{none}
\renewcommand\footnotetextcopyrightpermission[1]{}

\usepackage{listings}

\lstdefinelanguage{example}{
  keywords={handler, handle, perform, for, if, then, else, case, of},
  identifierstyle=\color{black},
  morecomment=[l]{--},
  morecomment=[l]{//},
  morestring=[b]",
  xleftmargin  = 2mm,
  morestring=[b]'
}

\lstdefinestyle{lambdap}{
    language=example,
  basicstyle=\small\ttfamily,
  keywordstyle=\sffamily\bfseries,
  captionpos=none,
  columns=flexible,
  keepspaces=true,
  showspaces=false,               
  showstringspaces=false,         
  showtabs=false,                 
  breaklines=true,                
  breakatwhitespace=true,         
  escapeinside={(*}{*)},
  literate={lam}{{$\lambda$}}1  {->}{{$\rightarrow$}}1 {Top}{{$\top$}}1 {o+o}{{$\oplus$}}1 {=>}{{$\Rightarrow$}}1 {/\\}{{$\Lambda$}}1 {<-}{{$\leftarrow$}}1  {[}{{$\langle$}}1   {]}{{$\rangle$}}1 {|->}{{$\mapsto$}}1  {\\}{{$\lambda$}}1,
  tabsize=2,
  commentstyle=\color{purple}\ttfamily,
  stringstyle=\color{red}\ttfamily,
  sensitive=false
}

\lstdefinestyle{haskell}{%
  language={},
  basicstyle=\small\ttfamily,
  keywordstyle=\sffamily\bfseries,
  captionpos=none,
  columns=flexible,
  keepspaces=true,
  showspaces=false,               
  showstringspaces=false,         
  showtabs=false,                 
  breaklines=true,                
  breakatwhitespace=true,         
  tabsize=2,
  commentstyle=\color{purple}\ttfamily,
  stringstyle=\color{red}\ttfamily,
  sensitive=false,
  morekeywords={type,data,infixr,class,where,instance,let,in,case,of},
}

\usepackage{cleveref}
\usepackage{mathpartir}
\usepackage{multicol}
 \usepackage{xspace}

\renewcommand{\for}{\text{\textbf{for}}}
\newcommand{\handler}{\text{\textbf{handler}}}
\newcommand{\perform}{\text{\textbf{perform}}}
\newcommand{\handle}{\text{\textbf{handle}}}
\newcommand{\tableval}[1]{\langle{#1}\rangle}
\newcommand{\return}{\mathit{return}}
\newcommand{\op}{\mathit{op}}
\newcommand{\traverse}{\mathit{traverse}}

\newcommand{\rulename}[1]{(\mathit{#1})}

\newcommand{\calculus}{$\lambda^{p}$\xspace}

\begin{document}

\title{Parallel Algebraic Effect Handlers}

\author{Ningning Xie}
\authornote{Both authors contributed equally to this research.}
\affiliation{%
  \institution{University of Cambridge}
  \country{United Kingdom}
}
\email{xnningxie@gmail.com}

\author{Daniel D. Johnson}
\authornotemark[1]
\affiliation{%
  \institution{Google Research}
  \country{Canada}
}
\email{ddjohnson@google.com}
    
\author{Dougal Maclaurin}
\affiliation{%
  \institution{Google Research}
  \country{USA}
}
\email{dougalm@google.com}

\author{Adam Paszke}
\affiliation{%
  \institution{Google Research}
  \country{Poland}
}
\email{apaszke@google.com}

\begin{abstract}
    Algebraic effects and handlers support
    composable and structured control-flow abstraction. However, existing designs of
    algebraic effects often require effects to be executed sequentially. This paper studies parallel algebraic
    effect handlers. In particular, 
    we 
    formalize \calculus, an untyped lambda calculus which models two key features,
    effect handlers and parallelizable computations, the latter of which takes the form of a \lstinline[language=example]{for}
    expression as inspired by the Dex programming language.
    We present various interesting examples expressible in our calculus,
    and provide a Haskell implementation.
    We hope this paper provides a basis for future designs and implementations of parallel algebraic effect handlers.
\end{abstract}

\maketitle

\section{Introduction}

Algebraic effects and handlers~\cite{Plotkin:effects,Pretnar:handlers} allow
programmers to define structured control-flow abstraction in a flexible and composable way.
Since introduced, they have been studied extensively in the community, supported in languages
including Koka~\cite{Leijen:koka}, Eff~\cite{Pretnar:introeff}, Frank~\cite{Lindley:Frank}
and Links~\cite{Lindley:effects}.
Recent work has implemented effect handlers
in Multicore OCaml~\cite{multicore:ocaml} to support asynchronous I/O for concurrent programming.

As an example of effect handlers, consider the monadic encoding of
the \lstinline[style=lambdap]{state} effect,
using the syntax of the untyped algebraic effect lambda calculus~\cite{Xie:evidently}:

\begin{lstlisting}[style=lambdap]
handler { get -> \x.\k. (\y. k y y)
        , set -> \x.\k. (\y. k () x) }
(\_. perform set 21; 
     x <- perform get (); 
     (\z. x + x)) 0 // 42
\end{lstlisting}
Here a \lstinline[style=lambdap]{handler} takes a list of operation clauses, and a computation
to be handled, which is represented as a unit-taking function. 
Inside each clause, \lstinline[style=lambdap]{x} is the argument to the operation,
\lstinline[style=lambdap]{k} is the \textit{resumption} captured by the handler, and each operation returns a
function, where \lstinline[style=lambdap]{y} is the monadic state that is threaded through computations.
Within the computation,
we \lstinline[style=lambdap]{perform} a set operation\footnote{For clarity,
we use \lstinline[style=lambdap]{x <- e1; e2} as a shorthand for \lstinline[style=lambdap]{(lamx. e2) e1}, and \lstinline[style=lambdap]{e1; e2}
for \lstinline[style=lambdap]{(lam_. e2) e1}.}, setting
the state to \lstinline[style=lambdap]{21}, and then \lstinline[style=lambdap]{get} the state, double it and return the result.
The initial state \lstinline[style=lambdap]{z} is set to \lstinline[style=lambdap]{0}. Evaluating the program gets us the result \lstinline[style=lambdap]{42}.

From this example, it may seem that algebraic effects, just like \textit{monads}~\cite{wadler1995monads},
generally need to be executed \textit{sequentially}. Indeed, in this case, \lstinline[style=lambdap]{perform set 21}
must be executed before \lstinline[style=lambdap]{perform get ()}, or otherwise we could get the state wrong.
Yet, recent work on Dex~\cite{dex},
a strict functional programming language for array programming,
has shown that it is \textit{possible} and \textit{useful}
to define \textit{parallel effect handlers}.
Specifically,
Dex supports a built-in effect \lstinline[style=lambdap]{Accum}
which works similarly to the \lstinline[style=lambdap]{State} effect, but can only be updated through an (infix) increment operation \lstinline[style=lambdap]{(+=)} and is implicitly initialized with an identity element of the increment.
\lstinline[style=lambdap]{Accum} is handled by the built-in handler \lstinline[style=lambdap]{runAccum}.
We can write the following program in Dex that sums up an input array
(we will introduce the syntax of Dex in \Cref{sec:overview:background}):

\begin{lstlisting}[style=lambdap]
sum = \x:(n=>Int).
  (_, total) = runAccum \y.
    for i:n. y += x.i
  total
\end{lstlisting}
Importantly, in this case, \lstinline[style=lambdap]{runAccum}
is able to run the "loop" (introduced by
the \lstinline[style=lambdap]{for} construct) in parallel!
The key reason \lstinline[style=lambdap]{runAccum}
can be run in parallel is that in Dex (1) updates
(\lstinline[style=lambdap]{+=}) to the accumulator reference (\lstinline[style=lambdap]{y}) must be \textit{additive contributions} over an associative operator,
opening up the parallelism by a potential reassociation of reduction steps;
and (2) there is no "read" operation in \lstinline[style=lambdap]{Accum} so
the state cannot be retrieved until \lstinline[style=lambdap]{runAccum} is complete,
making sure the increment in one iteration does not affect other iterations.


However, Dex only supports a few primitive effect handlers with built-in parallelization techniques.
So the key questions we ask in this paper are: \emph{is it possible to
support user-defined algebraic effects that preserve parallelism?
If so, what are their semantics}?

We offer the following contributions:

\begin{itemize}
    \item We illustrate the interaction between algebraic effects and paralleliable computations
    that take the form of the $\for$ construct as
    in Dex (\Cref{sec:overview:background}).
    \item We formalize \calculus, an untyped lambda calculus with parallel effect handlers (\Cref{sec:lambdap}).  
    Key to the design is a novel \textit{traverse} clause that allows user-defined behaviors around parallel regions.
    \item We present various interesting examples enabled by our design (\Cref{sec:examples}).
    \item We provide a Haskell implementation that captures the essence of \calculus.
    All examples presented in \Cref{sec:examples} can be encoded in the Haskell implementation.
    \item Finally, we discuss the challenges posed by our framework and future extensions that we hope to incorporate (\Cref{sec:discussion}).
\end{itemize}

\section{Background: Dex}
\label{sec:overview:background}

In this section, we present a brief overview of the Dex
programming language, and we refer the reader to
\cite{dex} for more detailed explanations.

Dex is a new array programming language with safe, efficient,
and differentiable typed indexing. Below presents a Dex example
that increments the elements in an array:
\begin{lstlisting}[style=lambdap]
incr = \x:(n=>Int). for i:n. x.i + 1
\end{lstlisting}
Here, \lstinline[style=lambdap]{x} of type \lstinline[style=lambdap]{n=>Int}
is an array indexed by indices of type \lstinline[style=lambdap]{n} and containing elements of type \lstinline[style=lambdap]{Int}.
Retrieval of individual elements is possible using the \lstinline[style=lambdap]{x.i} expression, which looks up an element of array corresponding to the index \lstinline[style=lambdap]{i}.
The construct \lstinline[style=lambdap]{for} builds an array, by repeatedly evaluating the body over the full range of the index type, which has to be finite.
In this case, the result array has the same length
as \lstinline[style=lambdap]{x}, with every element increased by 1.

What is important in the above example is that Dex is able to run the \lstinline[style=lambdap]{for} loop in parallel,
since the loop iterations do not depend on each other.
As argued by \citet{dex}, the \lstinline[style=lambdap]{for} expression enables a natural programming style for numerical computing, while being regular enough to enable compilation to efficient dense data-parallel code that can be later executed on hardware accelerators such as GPUs.

However, \lstinline[style=lambdap]{for} on its own has very weak expressive power.
With \lstinline[style=lambdap]{for} and array indexing alone, only pure maps over arrays can be expressed.
What makes the \lstinline[style=lambdap]{for} expression really useful is the ability to have effects in its body.
For example, the built-in \lstinline[style=lambdap]{State} effect makes it possible for a single loop iteration to influence evaluation of all subsequent iterations by modifying a value pointed to by a reference.
Unfortunately, while achieving great generality, \lstinline[style=lambdap]{State} has a significant drawback --- \lstinline[style=lambdap]{for} expressions with
the \lstinline[style=lambdap]{State} effect cannot be evaluated in parallel anymore.
To resolve this issue, Dex additionally implements a \emph{parallelism-friendly} \lstinline[style=lambdap]{Accum} effect.
It supports only a subset of operations that \lstinline[style=lambdap]{State} can express,
but it comes with the benefit of being parallelizable.
\lstinline[style=lambdap]{Accum} itself extends the power of \lstinline[style=lambdap]{for} expression to array reductions.
As an example,
let us revisit the program from the introduction:
\begin{lstlisting}[style=lambdap]
sum = \x:(n=>Int).
  (_, total) = runAccum \y.
    for i:n. y += x.i
  total
\end{lstlisting}
In this example, \lstinline[style=lambdap]{runAccum} takes a reference \lstinline[style=lambdap]{y}, which
is implicitly instantiated with \lstinline[style=lambdap]{0} (the identity element of \lstinline[style=lambdap]{Int} under addition),
and then builds
up an array, where each iteration calls 
the infix effectful operation \lstinline[style=lambdap]{(+=)} that adds
\lstinline[style=lambdap]{x.i} to \lstinline[style=lambdap]{y} and returns a unit.
Therefore, the \lstinline[style=lambdap]{for} builds an array of unit.
The \lstinline[style=lambdap]{runAccum} handles the operation, returning both the array (ignored by \lstinline[style=lambdap]{_}),
and the final reference value \lstinline[style=lambdap]{total}, which is returned as the final result.

Unfortunately,
while Dex can execute the \lstinline[style=lambdap]{for} loop inside \lstinline[style=lambdap]{runAccum} in parallel,
at the moment Dex lacks an extensible effect system,
and so its users are limited to a set of built-in implementations that the compiler can understand and compile using type-driven analysis to find the data-parallel regions in user programs.
This is highly unsatisfactory, and we outline many effects interesting for numerical computing that are unsupported by Dex today in \Cref{sec:examples}.



\begin{figure*}
\begin{tabular}{llllllll} \toprule
     expressions & $e$ & $\Coloneqq$ & $v \mid  e~e \mid \for~x:n.~e \mid$ $\handle~h~s~e $ & \\
     values      & $v$, $f$, $n$, $s$  & $\Coloneqq$ & $i \mid x \mid \lambda x.~e  \mid  \tableval{v_0,\dots,v_n} \mid \perform~\op  $\\
     handlers & $h$ & $\Coloneqq$ & $\{ \return \mapsto f_r, \op \mapsto f_p, \traverse \mapsto f_t  \}$ \\[5pt]
     evaluation context & $F$ & $\Coloneqq$ & $\square \mid F~e \mid v~F$ \\
                        & $E$ & $\Coloneqq$ & $\square \mid E~e \mid v~E \mid \handle~h~s~E$ \\
     \bottomrule
\end{tabular}

\vspace{10pt}

\begin{tabular}{lllll}
    $\rulename{app}$      & $(\lambda x.~ e)~v$ & $\longrightarrow$ &  $e[x:=v]$  \\
    $\rulename{index}$      & $\tableval{v_0,\dots,v_n}~ i$ & $\longrightarrow$ &  $v_i$  \\
    $\rulename{return}$   & $\handle~h~s~v$     & $\longrightarrow$ & $f_r~s~v$  & if $(\return \mapsto f_r) \in h$ \\
    $\rulename{perform}$  & $\handle~h~s ~E[\perform~\op~v]$ &  $\longrightarrow$ & $f_p~s~v~k$ & if $\op \notin \mathsf{bop}(E) \land (\op \mapsto f_p) \in h $ \\
    & & & & ~~where $k = \lambda s. \lambda x. ~ \handle~h~s~E[x]$\\
    $\rulename{traverse}$  & $\handle~h~s ~F[ \for~x:n.~e]$ &  $\longrightarrow$ & $f_t~n~s~\ell~k$ & if $(\traverse \mapsto f_t) \in h $ \\
    & & & & ~~where $\ell = \lambda ss. ~ \for~x:n.~\handle~h~(ss~x)~e$\\
    & & & & ~~\phantom{where} $k = \lambda s.~ \lambda xs. ~ \handle~h~s~F[xs]$\\
\end{tabular}

\begin{mathpar}
\inferrule*[right=$\rulename{step}$]{e \longrightarrow e'}{E[e] \longmapsto E[e']}
\and
\inferrule*[right=$\rulename{parallel}$]{\forall~0\le i < n.~e[x := i] \longmapsto v_i}{F[\for~x:n.~e] \longmapsto F[\tableval{v_0,\dots,v_{n-1}}] }
\end{mathpar}

\caption{Syntax and semantics of \calculus.}
\label{fig:syntax}
\end{figure*} 

\section{Parallel Effect Handlers}
\label{sec:lambdap}

In this section we introduce a calculus \calculus that lays out a basis for user-extensible
parallel effects in Dex and should provide a guidance for future implementation.
The syntax and semantics of \calculus are summarized in \Cref{fig:syntax}.

\subsection{Syntax}

Expressions $e$ include values $v$, applications $e~e$, the
$\for~x:n.~e$ construct to build an array of length $n$,
and the handle frame $\handle~h~s~e$,
which is a \textit{parameterized handler}~\cite{Pretnar:handlers}
that takes a handler $h$, a local state $s$,
and a computation $e$ to be handled.
Values $v$ include literals $i$, variables $x$, lambdas $\lambda x.~e$,
arrays $\tableval{v_0,\dots,v_n}$, and $\perform~\op$ that performs an operation.
We often use $f$ for lambdas, $n$ for literals, and $s$ for state.
While not included in the grammar, a $\handler~h~s~e$ construct that takes
a suspended computation $e$ to be handled can be defined as a syntactic sugar:
\begin{mathpar}
\handler~h~s~e \triangleq \lambda \_.~\handle~h~s~(e~())
\end{mathpar}

A handler $h$ defines the semantics of effects, where
for simplicity we assume that every effect has exactly one operation.
A handler takes three clauses:
(1) $\return \mapsto f_r$, a return clause that gets applied when the computation returns a value;
(2) $\op \mapsto f_p$, an operation clause that defines the operation implementation;
and (3) $\traverse \mapsto f_t $, a novel traverse clause critical to our calculus that handles parallel effects.
We discuss each clause in detail in the next section.

Evaluation contexts, essentially an expression with a hole ($\square$) in it,
explicitly indicate the evaluation order of an expression.
We distinguish between evaluation contexts $E$ and
\textit{pure} evaluation context $F$ that contains no $\handle$ frame.
The notation $E[e]$
denotes an expression obtained by substituting $e$ into the hole of $E$,
e.g., $((v~\square)~f)[e] = (v~e)~f$.

\subsection{Operational Semantics}

The bottom of \Cref{fig:syntax} defines the operational semantics of \calculus.
The evaluation rules have two forms: $\longrightarrow$ defines a primitive evaluation step,
and $\longmapsto$ evaluates expressions inside evaluation contexts.

\paragraph{Primitive evaluation rules}

We first discuss primitive evaluation rules.
Rule $\rulename{app}$ defines the standard call-by-value $\beta$-reduction.
Rule $\rulename{index}$ models the indexing operation in Dex as an application:
applying an array  $\tableval{v_0,\dots,v_n}$ to a literal $i$ projects out
the $i$th element $v_i$ in the array.

Rule $\rulename{return}$ and $\rulename{perform}$ define the original standard dynamic semantics
of effect handlers.
In particular, when a handler handles a computation, there are two possibilities.
If
the computation returns a value, then rule $\rulename{return}$
applies the return clause $f_r$ to the value. This is useful to model, for example, exceptions,
where an operation may cause the whole computation to return $\mathsf{Nothing}$,
while $f_r$ can be $\mathsf{Just}$ that wraps the value when the computation returns normally.
If the computation performs an operation
$\perform~\op~v$ that calls the operation $\op$ with the argument $v$,
then
rule $\rulename{perform}$
finds the innermost handler for the operation (specified as $\op \notin \mathsf{bop}(E)$),
and applies the operation clause $f_p$ to the state $s$, the operation argument $v$,
as well as the resumption $k$.
The resumption $k$ takes a new handler state $s$ and the operation result $x$,
and captures the handler with the new state and the evaluation context between the handler
and the operation call.

\paragraph{Traverse}

Rule $\rulename{traverse}$ captures the essence of parallel effect handlers in \calculus,
adding a third option of how the computation to be handled can interact with the handlers.
Specifically, if the computation calls $\for~x:n.~e$, then we would like the expression $e$ to be
executed in parallel for each $x$ in $n$. However, naively evaluating $e$ could get us stuck,
as the expression may perform operations!
Instead, we allow the users to define how a $\for$ expression should be handled.
In particular,
rule $\rulename{traverse}$ first finds the innermost handler $h$,
and applies its traverse clause $f_t$ to (1) the array length $n$,
(2) the new state $s$, (3) the $\for$ expression $\ell$, and (4) and resumption $k$ that resumes the program segment following the loop.

There are several things to be noted here.
First, $h$ is the innermost handler for any operation rather than for a specific operation.
The difference here from rule $\rulename{perform}$ can be seen from the use of $F$ (instead of $E$)
when looking for handlers. One way to interpret the rule is that $\for$ is an effect
that can be handled by any handler -- this is true in the formalism as every handler defines
the traverse clause (we discuss sequential effect handlers in \Cref{sec:discussion:sequential}).
Second, $\ell$
wraps the original $\for$ expression, but pushes the handler inside the $\for$
expression, and thus the corresponding operations in $e$ can be handled by $h$.
Third, we need to update the state for the handlers in $\ell$. To this end,
$\ell$ takes an array of states $ss$, and during each iteration
the handler takes the corresponding
state from the array by indexing $(ss~x)$.

Now depending on the implementation of $f_t$,
the program can have different behaviors.
(i) $f_t$ may never call $\ell$. Then the whole computation of the $\for$ expression is discarded.
(ii) $f_t$ may call $\ell$ exactly once. Then the $\for$ expression will keep propagating
to outer handlers. When there is no outer handler, it means all handlers have properly
handled the $\for$ expression, and thus we are able to
execute the expressions in parallel (in rule $\rulename{parallel}$, which we discuss shortly).
(iii) $f_t$ may call $\ell$ multiple times, then the same $\for$ expression will be evaluated multiple times.
We consider some examples of
different behaviors. For instance,
if the 
handler has no special behavior for parallelism,
a default implementation of the traverse clause can be (case (ii)):
\begin{mathpar}
\traverse \mapsto \lambda n.~ \lambda s.~ \lambda \ell.~\lambda k.~k~s~(\ell~(\for~x:n.~s))
\end{mathpar}
The handler may also just ignore $\ell$ and pass something totally different to $k$ (case (i)):
\begin{mathpar}
\traverse \mapsto \lambda n.~ \lambda s.~\lambda \ell.~\lambda k.~k~s~\tableval{1,2,3}
\end{mathpar}
Note how this corresponds nicely to how handlers handle resumptions:
a resumption may never be called (e.g., for exception handlers),
or called exactly once (for most handlers including, e.g., reader),
or called multiple times (for non-determinisim).

\paragraph{Evaluation inside evaluation contexts.}

Now we turn to the rules of $\longmapsto$, which evaluates expressions inside evaluation contexts.
Rule $\rulename{step}$ says that if an expression $e$ can take a primitive evaluation step to $e'$,
then the whole expression $E[e]$ evaluates to $E[e']$.

Rule $\rulename{parallel}$ is where parallelism takes place.
Specifically,
when we have a $\for$ expression not under any handlers (recall that $F$ is a pure evaluation context),
it means all handlers have been pushed inside the $\for$ expression, and so we are ready to evaluate the
$\for$ body in parallel! For every $i$ ranging from $0$ up to $n$, we evaluate
the expression $e$ after substituting $x$ by $i$.
Here we assume a built-in parallelism support for evaluating the $\forall$ parallelism (which can be,
for example, the built-in parallelism support for $\for$ in Dex).

\paragraph{Example}

Now let us consider a paralleliable reader handler as an example.
For the sake of readability, in the example we ignore handler states, and let $\ell$ take a unit.

\begin{tabular}{lll}
$h$ & $=$  & $\{~ return \mapsto \lambda x.~x,~ ask \mapsto \lambda x.~\lambda k.~k~42$\\
  & & $ ,~ traverse \mapsto \lambda n.~\lambda \ell.~\lambda k.~ k~( \ell ~() )~\}$\\
$\ell$ & $=$ & $\lambda\_.~\for~x:5.~\handle~h~(\perform~ask~()) $\\
$k$ & $=$ & $\lambda xs.~ \handle ~h~xs $ \\
\end{tabular}

\noindent Now we have (we use $\longmapsto^{*}$ as the transitive closure of $\longmapsto$):

{
\raggedright
\begin{tabular}{ll@{\hskip -4pt}r}
& $\handle~h~(\for~x : 5.~\perform~ask~())$ \\
$\longmapsto^{*}$ 
& $ (\lambda n.~\lambda \ell.~\lambda k.~ k~( \ell~()  ))~5~\ell~k $
& $\rulename{traverse}$ \\
$\longmapsto^{*}$
& $ k~( \ell~()  )$
& $\rulename{app}$ \\
$\longmapsto^{*}$
& $ k~( \for~x:5.~\handle~h~(\perform~ask~())  )$ 
& $\rulename{app}$ \\
$\longmapsto^{*}$
& $ k~ \tableval{42,42,42,42,42}  $  
& $\rulename{parallel}$
\\
$\longmapsto^{*}$
& $ (\lambda x.~x)~\tableval{42,42,42,42,42}  $  
& $\rulename{return}$
\\
$\longmapsto^{*}$
& $ \tableval{42,42,42,42,42}  $  
& $\rulename{app}$
\\
\end{tabular}
}
See also Appendix \ref{app:accum-reduction} for a more sophisticated reduction that uses handler states to implement our Accum effect.

%

\section{Examples}
\label{sec:examples}
Now that we have described our system, in this section we will show how we can implement
a variety of interesting and useful effects. We will express these examples using a richer surface language that includes tuples, conditionals, infix operators, algebraic data types, numbers, and strings.

\subsection{Accum}\label{subsec:examples:accum}
We begin by showing how to express the parallel accumulation effect in our language. To handle the effect, we must provide 
an associative binary operation \texttt{(<>)} and an identity for that operation (essentially forming a \textit{monoid}). For instance, to sum an array of numbers, we can use:
\begin{lstlisting}[style=lambdap]
sum = (\xs.
  (_, total) <- runAccum (+) 0 (\_.
    for i:(length xs). perform accum (xs i));
  total)
\end{lstlisting}

Because the binary operator is associative, we are free to independently compute results for each list iteration, and then combine them in our traverse handler.
\begin{lstlisting}[style=lambdap]
runAccum = \(<>). \mempty. \f.
  handle { return |-> \s.\x. (x, s),
           accum |-> \s.\x.\k. k (s <> x) (),
           traverse |-> (\n.\s.\l.\k.
              pairs <- l (for i:n. mempty);
              results <- for i:n. (fst (pairs i));
              outs <- for i:n. (snd (pairs i));
              out <- reduce (<>) outs;
              k (s <> out) results)
         } mempty (f ())
\end{lstlisting}
Here we assume the existence of a parallelizable function \texttt{reduce} which reduces a table into a single value by applying \texttt{(<>)}. Due to space limitations we do not implement \texttt{reduce} here; roughly, it corresponds to a parallel reduction circuit of depth $O(\log n)$ constructed by forming a balanced binary tree over array elements and applying \texttt{(<>)} at each node.

\subsection{Weak Exceptions}
Our effect system can also express a form of exception handling. However, since loop iterations are always evaluated in parallel, these exceptions are "weak": an exception in one iteration of a loop does not interrupt execution in any other iterations, although it will still prevent execution of the code after the loop body. Our handler has the following form:
\begin{lstlisting}[style=lambdap]
data Either a b = Left a | Right b
runWeakExcept = \f.
  handle { return |-> \_.\x. Right x,
           throw |-> \_.\err.\k. Left err,
           traverse |-> (\n.\_.\l.\k.
              eithers <- l (for i:n. ())
                combined <- firstFailure eithers
                case combined of
                  Left err -> Left err
                  Right res -> k () res
          } () (f ())
\end{lstlisting}
Here \texttt{firstFailure} takes a table of Eithers and returns either the first Left, or the table of values if all values were wrapped in Right; it can be implemented in terms of \texttt{reduce}.

We can observe the ``weak" nature of these exceptions by combining them with another effect:
\begin{lstlisting}[style=lambdap]
(res, out) = runAccum (++) "" (\_. 
  runWeakExcept (\_.
    perform accum "start ";
    for i:5. if i == 2
      then (perform accum "!";
            perform throw "error";
            perform accum "unreachable")
      else (perform accum (toString i));
    perform accum " end")
  // (Left "error", "start 01!34")
\end{lstlisting}
In this example, all loop iterations execute their effects in parallel, and then computation aborts at the end of the for loop.
(Here \texttt{(++)} is the string concatenation operator.)

\subsection{PRNG}
One effect that is particularly useful for real-world numerical computation is the generation of (pseudo)random numbers. However, doing so in a parallelizable way is nontrivial. 
Suppose we want to run a computation like this in parallel:
\begin{lstlisting}[style=lambdap]
binomial_times_uniform = \n. \p.
  (_, count) = runAccum (+) 0 (\_.
    for _:n.
      u <- perform sampleUniform ();
      if u < p then (perform accum 1) else ()))
  v <- perform sampleUniform ();
  count * v
\end{lstlisting}
This example computes a binomial random variable by summing a collection of weighted coin flips, then scales it by another random variable, and we want each coin flip to draw distinct random numbers, but also execute in parallel.

One way to handle this is using a ``splittable PRNG" \citep{claessen2013splittable}, whose state (called a ``key'') can be split into arbitrarily many independent streams of random numbers; this technique is used to implement accelerator-friendly random numbers in the library JAX \citep{jax_prng}. Conveniently, this design can be directly mapped to our parallel effects system.
We assume the existence of two functions: \texttt{splitKey}, which takes a key and a natural number, and returns a table of new keys; and \texttt{sampleUniform}, which takes a key and returns a random number between 0 and 1. Given this, we can implement a simple random number effect as follows:
\begin{lstlisting}[style=lambdap]
runRandom = \seed. \f.
  handle { return |-> \key.\x. x,
           sampleUniform |-> (\key.\_.\k.
              [key1, key2] <- splitKey key 2;
              u <- genUniform key1;
              k key2 u),
           traverse |-> (\n.\key.\l.\k.
              [key1, key2] <- splitKey key 2;
              results <- l (splitKey key1 n);
              k key2 results)
         } seed (f ())
\end{lstlisting}
Here we transform effectful expressions into functions from a PRNG key to a value. We handle \texttt{sampleUniform} by splitting the key, then using one result to generate the uniform and the other to run the continuation. We handle $\for$ loops similarly, except that the first key is split again to generate independent streams of random numbers for each loop iteration.

An interesting property of this handler is that the following computations may have different results:
\begin{lstlisting}[style=lambdap]
result_1 = runRandom shared_seed (\_.
  u0 <- perform sampleUniform ();
  u1 <- perform sampleUniform ();
  u2 <- perform sampleUniform ();
  [u0, u1, u2])
  
result_2 = runRandom shared_seed (\_.
  for i:3. perform sampleUniform ())
\end{lstlisting}
This highlights an important property of the \for{} construct in \calculus, in contrast to similar looping constructs in other languages: the semantics of a program with a \for{} expression is not necessarily equivalent to the semantics of a program with a sequentially-unrolled loop in its place.

\subsection{Amb}
Our final example is the \texttt{Amb} effect \citep{mccarthy1961basis} (also known as the list monad). Conceptually, the \texttt{amb} operator takes as argument a table of values, and nondeterministically picks one. Unlike the PRNG effect, however, the result of a computation in the \texttt{Amb} is not a single result but instead the table of \textit{all} possible results we might obtain:
\begin{lstlisting}[style=lambdap]
result = runAmb (\_.
  chars <- for i:3. perform amb ["H", "T"]);
  reduce (++) "" chars)
// result == ["HHH", "HHT", "HTH", "HTT",
//            "THH", "THT", "TTH", "TTT"]
\end{lstlisting}

Unlike the other effect handlers we have introduced, the handler for $\texttt{amb}$ can introduce parallelism into sequential code, by calling the continuation inside a parallel loop:
\begin{lstlisting}[style=lambdap]
runAmb = \f.
  handle { return |-> \_.\x. [x],
            amb |-> (\_.\options.\k.
              n <- length options;
              for i:n. k () (options i)),
            traverse |-> (\n.\_.\l.\k.
                results <- l (for i:n. ())
                productElts <- cartesianProd results
                n <- length productElts;
                for i:n. k () (productElts i)),
          } () (f ())
\end{lstlisting}
Here \texttt{cartesianProd} is a function which takes a length-$m$ table of variable-length tables, and returns a variable-length (called $n$) table of length $m$ tables, such that each element of the result is formed by taking one element from each of the original variable-length tables.

Due to the compositionality of our system, users are free to nest multiple effects. For instance, by nesting \texttt{runAmb} inside \texttt{runAccum}, we can count samples with certain properties:
\begin{lstlisting}[style=lambdap]
// How many pairs of single-digit numbers add up
// to 13?
(_, count) = runAccum (+) 0 (\_. runAmb (\_.
  d1 <- perform amb [0,1,2,3,4,5,6,7,8,9];
  d2 <- perform amb [0,1,2,3,4,5,6,7,8,9];
  if (d1 + d2 == 13)
    then perform accum 1
    else () ))
\end{lstlisting}
Let us emphasize again that even though the code example looks entirely serial, it will be converted into a parallel loop over all valid values for \lstinline[style=lambdap]{d1} and \lstinline[style=lambdap]{d2} by the \lstinline[style=lambdap]{amb} effect.

\section{Discussion}
\label{sec:discussion}

In this section we first outline the Haskell implementation of \calculus,
and then discuss potential limitations and future directions for the work outlined in this text.

\subsection{Haskell implementation}
\label{sec:haskell}
We provide an implementation of our parallel effect system in Haskell, along with Haskell versions of each of the examples described in Section \ref{sec:examples}. For ease of implementation and use, we build our implementation on top of Haskell's own interpreter and inherit its execution semantics. As such, the implementation may not actually run loop iterations in parallel, and reductions may not be applied in the same order as specified in Figure \ref{fig:syntax}. Nevertheless, it exhibits the same overall behavior as \calculus,
including in particular the interaction between \lstinline[style=lambdap]{for}
and \lstinline[style=lambdap]{traverse}.

Our Haskell implementation is based on the Free monad and the FEFree monad as described by \citet{kiselyov2015freer}, which represents an effectful computation by capturing the first performed effect (the equivalent of $E[\perform~\op~v]$ in \calculus) and storing it as a pair of a pure value (e.g. $v$) and continuation ($\lambda x.~E[x]$). We introduce a Haskell type for arrays, along with a special function $\for$ to construct them. We then construct a type for parallel effectful computations (\texttt{EffComp effectRow a}) which captures either the first effect (as in FEFree) or the first for loop (the equivalent of $F[ \for~x:n.~e]$ in \calculus), storing the latter as a pair of loop iteration function $\lambda x.~e$ and continuation $\lambda ys.~F[ys]$. We then implement the reduction rules for \textit{(perform)}, \textit{(traverse)}, and \textit{(parallel)} by recursively pattern-matching on this data type. See \cref{app:haskell-impl} for more details, and the supplemental material for the implementation itself.

Although \calculus is untyped, the Haskell implementation suggests one way to formalize the type system of parallel algebraic effects. One interesting feature is that $\return$ (along with $\perform$ and $\traverse$) is polymorphic in its output type, since the handler may be locally applied to the body of multiple $\for$ expressions with different types. If a handler needs to change the return type, it must do so by wrapping the polymorphic return type \texttt{a} with a functor-like higher-order type \texttt{f a}. We leave the details of a typed version of \calculus to future work.

\subsection{Sequential algebraic effects}
\label{sec:discussion:sequential}

So far, our focus was the design of a formalism for an extensible \emph{parallel} effect system.
While the final result is sufficiently powerful to model a Dex-like accumulation effect and a wide range of other examples, it does fall short of modeling effects that are \emph{sequential}, in the sense that no viable \lstinline[style=lambdap]{traverse} method exists for them.
One good example is the state effect, which requires the $\for$ iterations to be executed in order, since arbitrary stateful updates from earlier iterations have to be visible in the subsequent ones.
Dex implements a \lstinline[style=lambdap]{State} effect along with \lstinline[style=lambdap]{Accum}, and uses a type-directed compilation strategy to parallelize loops that are pure or only use associative accumulation, while the $\for$ expressions with bodies that have a \lstinline[style=lambdap]{State} effect are compiled as sequential loops.
We hope to address this limitation in future work, which should present a coherent effect system unifying the sequential and parallel execution strategies for effects, and hence model the design of Dex even more closely.

\subsection{Handlers returning functions}

One interesting property of the formalism presented here is that every handler has to be associated with a functorial data type, which follows from the fact that the traverse rule of \Cref{fig:syntax} constructs an array of elements of completely arbitrary type, by wrapping their computation in the same handler.
So far we have seen examples where this functor is well-behaved, but it turns out that it is not always the case.
In particular, if the handler returns the continuation itself without calling it (corresponding to the functor \texttt{(a $\rightarrow$)} for some \texttt{a}), it can break up a single parallel loop into two: one containing the effects outside each iteration's continuation and the other containing the effects inside them.
While initially this might seem relatively benign, this introduces synchronization points that can significantly alter the semantics of the program in quite surprising ways. 

In fact, this behavior is one of our main reasons for presenting a "stated" version of our effect system.
The usual trick of encoding state by making the handler return a result of type \lstinline[style=lambdap]{s -> (a, s)} does not work in our paradigm, because it forces an additional and unwanted synchronization of the $\for$ loop.
One way to limit this issue would be to forbid the functor type associated with handler to include the function type, effectively making the functions second-class from the point of the effect system.
To some extent this is already the case in Dex, where e.g. the type of values valid to be carried in a \lstinline[style=lambdap]{State} effect is restricted to never include functions, so as to enable elimination of higher-order functions (a specialized form of defunctionalization).

\section{Conclusion}

In summary,
we have designed a calculus \calculus for parallel effect handlers, where
paralleliable
$\for$ expressions are handled by the traverse clause
in handlers,
and eventually non-effectful $\for$ expressions can run in parallel.
As future work, we would like to investigate a typed formalism of \calculus
that can also express sequential algebraic effect handlers,
and implement \calculus as a source-to-source transformation in Dex.

\bibliographystyle{ACM-Reference-Format}
\bibliography{main}

\appendix

\section{Example reduction of accum effect}\label{app:accum-reduction}
Here we show the steps taken while reducing our accumulation example, according to the reduction rules of \calculus. Consider the following program and handler (reproduced from \cref{subsec:examples:accum}):

\begin{lstlisting}[style=lambdap]
runAccum = \(<>). \mempty. \f.
  handle { return |-> \s.\x. (x, s),
           accum |-> \s.\x.\k. k (s <> x) (),
           traverse |-> (\n.\s.\l.\k.
              pairs <- l (for i:n. mempty);
              results <- for i:n. (fst (pairs i));
              outs <- for i:n. (snd (pairs i));
              out <- reduce (<>) outs;
              k (s <> out) results)
         } mempty (f ())

sum = (\xs.
  (_, total) <- runAccum (+) 0 (\_.
    for i:(length xs). perform accum (xs i));
  total)
  
value = sum [1, 2, 3]
\end{lstlisting}
Abbreviating our handler as $h$, and following the reduction rules described in \cref{fig:syntax}, we obtain the following sequence:
\begin{lstlisting}[style=lambdap]
sum [1, 2, 3]
// (app)
(_, total) <- handle h 0 (
    for i:3. perform accum ([1, 2, 3] i));
total
// (traverse)
n <- 3;
s <- 0;
l <- \ss. for i:n. handle h (ss i) (
    perform accum ([1, 2, 3] i));
k <- \s. \x. handle h s x;
(_, total) <- (
    pairs <- l (for i:n. 0);
    results <- for i:n. (fst (pairs i));
    outs <- for i:n. (snd (pairs i));
    out <- reduce (+) outs;
    k (s + out) results);
total
// ...
pairs <- (\ss. for i:3. handle h (ss i) (
    perform accum ([1, 2, 3] i)) (for i:3. 0)
results <- for i:3. (fst (pairs i))
outs <- for i:3. (snd (pairs i))
out <- reduce (+) outs;
(_, total) <- (\s. \x. handle h s x)
    (0 + out) results;
total

// ...
pairs <- for i:3. handle h ((for i:3. 0) i) (
    perform accum ([1, 2, 3] i))
results <- for i:3. (fst (pairs i))
outs <- for i:3. (snd (pairs i))
out <- reduce (+) outs;
(_, total) <- (\s. \x. handle h s x)
    (0 + out) results;
total
\end{lstlisting}
At this point, the rule $\rulename{parallel}$ applies, so we independently reduce each iteration to a value. For $i = 0$:
\begin{lstlisting}[style=lambdap]
handle h ((for i:3. 0) 0) (
    perform accum ([1, 2, 3] 0))
// (parallel)
handle h ([0, 0, 0] 0) (perform accum ([1, 2, 3] 0))
// (index)
handle h 0 (perform accum 1)
// (perform)
s <- 0;
v <- 1;
k <- \x. \s. handle h s x;
k (s + v) ()
// ...
handle h 1 ()
// (return)
((), 1)
\end{lstlisting}
The other iterations reduce to $((), 2)$ and $((), 3)$ respectively. We then resume reducing the full program using $\rulename{step}$:
\begin{lstlisting}[style=lambdap]
// (parallel)
pairs <- [((), 1), ((), 2), ((), 3)];
results <- for i:3. (fst (pairs i))
outs <- for i:3. (snd (pairs i))
out <- reduce (+) outs;
(_, total) <- (\s. \x. handle h s x)
    (0 + out) results;
total
// ...
results <- [(), (), ()];
outs <- [1, 2, 3];
out <- reduce (+) outs;
(_, total) <- (\s. \x. handle h s x)
    (0 + out) results;
total
// ...
out <- 6;
(_, total) <- (\s. \x. handle h s x)
    (0 + out) [(), (), ()];
total
// ...
(_, total) <- handle h 6 [(), (), ()];
total

// (return)
(_, total) <- ([(), (), ()], 6);
total
// ...
6
\end{lstlisting}

\section{Overview of Haskell implementation}\label{app:haskell-impl}
Here we give a brief overview of the structure of the Haskell implementation. For all of the details, including Haskell versions and execution results for each of the example handlers and programs in \cref{sec:examples}, see the supplemental material.

We define a type for tables (arrays), using Haskell's type-level literals to annotate the table length (as an approximation of Dex's table arrow and \texttt{Fin n} type):
\begin{lstlisting}[style=haskell]
data Table (n :: Nat) a where
  UnsafeFromList :: forall n a. KnownNat n
                 => [a] -> Table n a
\end{lstlisting}

Following the FEFree monad \citep{kiselyov2015freer}, we assume that effects are functor-like higher kinded types parameterized by the result of each effect, e.g. for an effect \texttt{State s :: * -> *} we might have \texttt{Get :: State s s} and \texttt{Put :: s -> State s ()}. We additionally add a notion of lifting effects into ordered ``effect rows'':
\begin{lstlisting}[style=haskell]
type EffSig = * -> *
data EffCons (sig :: EffSig) (sigs :: EffSig) r = Here (sig r) | There (sigs r)
infixr 5 `EffCons`
data EffNil r  -- pure computation; no operations

class HasEff (sig :: EffSig) (row :: EffSig)
    where liftEff :: sig r -> row r

instance HasEff sig (sig `EffCons` rest)
    where liftEff op = Here op

instance HasEff sig rest
         => HasEff sig (other `EffCons` rest)
    where liftEff op = There (liftEff op)

\end{lstlisting}
For instance, \lstinline[style=haskell]{State s `EffCons` Except e `EffCons` EffNil} is the signature that states \texttt{State s} and \texttt{Except e} are both possible operations, and a value of type \lstinline[style=haskell]{(State s `EffCons` Except e `EffCons` EffNil) r} represents either a state operation or an except operation whose return type is \texttt{r}, tagged with which handler should handle it. \texttt{liftEff} then finds the handler that should handle a given effect. (For readability, we have omitted a helper type family that enables Haskell to avoid an overlapping instance error by always choosing the leftmost handler if possible.)

\pagebreak
Given a particular effect row, our parallelism-aware FEFree-like monad is defined as
\begin{lstlisting}[style=haskell]
data EffComp (sig :: EffSig) r where
  -- This computation is just a pure value.
  Pure :: r -> EffComp sig r
  -- This computation is equivalent to
  -- an E[perform op v] or F[for i:n. e]
  Impure :: EffOrTraversal sig r
         -> (r -> EffComp sig s)
         -> EffComp sig s
  
-- Helper which holds either an operation,
-- or a parallel loop.
data EffOrTraversal sig r where
  -- equivalent of "perform op v"
  Effect :: sig r -> EffOrTraversal sig r
  -- equivalent of "for i:n. e"
  TraverseTable :: Table n (EffComp sig s) -> EffOrTraversal sig (Table n s)
  
-- Sequencing computations using a monad
instance Monad (EffComp sig) where
  return = Pure
  (Pure v >>= f) = f v
  (Impure va vc >>= f) =
    Impure va $ \a -> vc a >>= f
\end{lstlisting}

We construct the Haskell equivalent of the \perform{} and \for{} expressions:
\begin{lstlisting}[style=haskell]
perform :: HasEff sig union
        => sig r -> EffComp union r
perform e = Impure (Effect $ liftEff e) Pure

iota :: forall n. KnownNat n => Table n Int
iota = let nv = fromIntegral $ natVal (Proxy @n)
       in UnsafeFromList @n [0 .. nv - 1]

traverseTable :: (a -> EffComp sig b)
              -> Table n a
              -> EffComp sig (Table n b)
traverseTable f a =
  Impure (TraverseTable (f <$> a)) Pure

for :: KnownNat n => (Int -> EffComp sig b)
                  -> EffComp sig (Table n b)
for = flip traverseTable iota
\end{lstlisting}

A handler is defined using a Haskell record containing the three functions required by the handler. The type of a handler is parameterized by the operations it handles, the remaining effects in the row, the type of state it carries, and a functor \texttt{f} such that, if the original computation produces an \texttt{a}, the handler produces an \texttt{f a}. (For instance, for \texttt{Except e} we have \texttt{f = Either e}, and thus th ehandler produces values of type \texttt{Either e a}.)
\begin{lstlisting}[style=haskell]
data ParallelizableHandler (op :: EffSig)
        (m :: * -> *) s (f :: * -> *) =
  ParallelizableHandler
  { handleReturn   :: forall a.
      s -> a -> m (f a)
  , handlePerform  :: forall a b.
      s -> op a -> (s -> a -> m (f b)) -> m (f b)
  , handleTraverse :: forall a b n. KnownNat n =>
      s -> (Table n s -> m (Table n (f a)))
        -> (s -> Table n a -> m (f b))
        -> m (f b)
  }
\end{lstlisting}

Finally, we provide functions to handle individual effects in an effect row, as well as to execute pure code. Our handler is represented as a Haskell record containing the three functions required by the handler.

\begin{lstlisting}[style=haskell]
runPure :: EffComp EffNil r -> r
runPure = \case
  Pure r -> r
  Impure (Effect eff) cont ->
    case eff of {} -- no operations, not possible
  Impure (TraverseTable iters) cont ->
    runPure $ cont $ fmap runPure iters

handle :: forall op rest s f a
    . ParallelizableHandler op (EffComp rest) s f
   -> s
   -> EffComp (op `EffCons` rest) a
   -> EffComp rest (f a)
handle h s comp = case comp of
    -- (return)
    Pure r -> (handleReturn h) s rv
    -- (perform)
    Impure (Effect (Here op)) cont ->
        (handlePerform h) s op (\s a ->
              handle h s $ cont a)
    -- (traverse)
    Impure (TraverseTable (iters :: Table n b))
           cont ->
        (handleTraverse h) s runIters runCont
    where
        runIters ss = for $ \i ->
            handle h (tableIndex ss i)
                     (tableIndex iters i)
        runCont s a = handle h s (cont a)
    -- ignore other operations
    Impure (Effect (There op)) cont ->
        Impure (Effect op) (handle h s . cont)
\end{lstlisting}

\end{document}